\documentclass[twocolumn,showpacs,preprintnumbers,amsmath,amssymb,floatfix]{revtex4-1}

\usepackage{multirow}
\usepackage{graphicx}
\usepackage{longtable}
\usepackage{dcolumn}
\usepackage{bm}
\usepackage{fancybox}
\usepackage{lineno}

\newcommand{\bfr}{{\bf r}}

\newcommand{\bfR}{{\bf R}}

\def\H0{H^0}

\newcommand{\req}[1]{\mbox{Eq.~\!(\ref{#1})}}

\def\connect#1{\leavevmode{\setbox1=\hbox{#1}\copy1%
\raise .2\ht1 \vbox{\moveleft \wd1\vbox{\hrule width \wd1 height .5pt depth 0pt}}%
}}

\def\ftn[#1]{\rlap{\footnotemark[#1]}}

\begin{document}
\title{
A finite electric-field approach to evaluate
the vertex correction for the screened Coulomb interaction
in the quasiparticle self-consistent GW method}
\author{Hirofumi Sakakibara}
\author{Takao Kotani}
\affiliation{Department of Applied Mathematics and Physics, Tottori university, Tottori 680-8552, Japan}
\author{Masao Obata}
\author{Tatsuki Oda}
\affiliation{Department of Computational Science, Institute of Science and Engineering, Kanazawa University,
Kakuma, Kanazawa 920-1192, Japan}

\date{\today}
\begin{abstract}
We apply the quasiparticle self-consistent GW method (QSGW) to 
slab models of ionic materials, LiF, KF, NaCl, MgO, and CaO, under electric field.
Then we obtain the optical dielectric constants
$\epsilon_\infty({\rm Slab})$ from the differences of the slopes of
the electrostatic potential in the bulk and vacuum regions.
Calculated $\epsilon_\infty({\rm Slab})$ show very good
agreements with experiments. For example,
we have  $\epsilon_\infty({\rm Slab})$=2.91 for MgO, in agreement with
the experimental value $\epsilon_\infty(\rm Experiment)$=2.96.
This is in contrast to $\epsilon_\infty(\rm RPA)$=2.37, which is calculated in 
the random-phase approximation for the bulk MgO in QSGW. 
After we explain the difference between the quasiparticle-based perturbation
theory and the Green's function based perturbation theory,
we interpret the large difference $\epsilon_\infty({\rm Slab})-\epsilon_\infty({\rm RPA})=2.91-2.37$
as the contribution from the vertex correction of the proper
polarization which determines the screened Coulomb interaction $W$.
Our result encourages the theoretical development of self-consistent
$G_0 W$ approximation along the line of QSGW self-consistency, as
was performed by Shishkin, Marsman and Kresse
[Phys. Rev. Lett. {\bf 99}, 246403(2007)].
 

%

\end{abstract}
\maketitle


\section{introduction}
The quasiparticle self-consistent GW (QSGW) is one of the most
reliable method to determine the one-particle effective Hamiltonian
which describes the independent-particle picture,
or the quasiparticle (QP) picture,
for treating electric excitations of materials \cite{Faleev04,vans06,kotani07a}. 
Other competitive methods such as HSE \cite{hse06} and 
Tran-Blaha-09 functional \cite{PhysRevLett.102.226401} 
may work well in many systems, although we may need to use
material-dependent parameters \cite{he_screened_2012}.
In contrast, QSGW is virtually parameter free and gives reliable descriptions for
a wide range of materials, not only metals and semiconductors, but also 
transition-metal oxides, type-II superlattice, and $4f$ systems
\cite{di_valentin_quasiparticle_2014,deguchi_accurate_2016,okumura_spin-wave_2019,lee_role_2020}.
Since heterogeneous mixtures of materials are used in current technologies,
QSGW is worth to be developed more as a tool to treat electronic
structures of such materials, where methods including such material-dependent parameters are hardly
applicable.

However, QSGW as it is has a shortcoming that it gives a systematic
overestimation of the exchange effects. This results in 
a little larger band gaps in QSGW for materials.
In fact, Faleev, van Schilfgaarge, and Kotani \cite{Faleev04,vans06,kotani07a}
made a suggestion that the overestimation is removed if we perform
improved QSGW calculations taking into account the enhancement
of the screening effect due to the electron-hole correlation
in the evaluation of the screened Coulomb interaction $W$.
This is based on the theoretical consideration combined with the observation
that the calculated optical dielectric constant $\epsilon_\infty$ in the
random-phase approximation (RPA) in
QSGW gives $\sim 20$ percent smaller $\epsilon_\infty$
than experiments for kinds of materials \cite{kotani07a,PhysRevMaterials.2.013807}.

Such an improved calculation was performed by Shishkin, Marsman and
Kresse, where they include the enhancement of the screening effect \cite{shishkin_accurate_2007}.
The enhancement is via the vertex correction for the proper polarization $P$, which
determines $W=v/(1-vP)$, where $v$ denoted the Coulomb interaction.
They approximately include the lowest-order vertex correction 
due to the electron-hole correlation; see Eq.(15) around in Ref.\onlinecite{Bruneval05}.
Their results are theoretically quite satisfactory in the sense
that both band gaps and $\epsilon_\infty$,
which are calculated simultaneously and self-consistently without
parameters as in HSE, are in agreement with experiments.
For example, calculated values $\epsilon_\infty=2.96$
and band gap $E_{\rm G}=8.12$eV for MgO are in agreement with the experiments, 2.95, and 7.83
eV, respectively. See sc$GW$(e-h) in Table I in Ref.\onlinecite{shishkin_accurate_2007}.
Furthermore, based on these theoretical analyses, we can introduce QSGW80
to avoid the very expensive computational costs of the method by
Shishkin {\it et al.}. QSGW80 is just a simple hybridization, 80 \% QSGW+ 20 \% GGA
to include such enhancement of the screening effectively.
The hybridization is very different from the hybridization in HSE,
where the mixing ratio $\alpha$ between GGA and the Hartree-Fock strongly affects to final results.
QSGW80 works well to describe experimental band gaps \cite{chantis06a}.
The performance of QSGW80 is systematically examined in
Ref.\onlinecite{deguchi_accurate_2016} by Deguchi {\it et al.},
where we see both the calculated band gaps and effective masses are in good
agreements with experiments. QSGW80 is successfully used for practical applications, for example,
to the type-II superlattice of InAs/GaSb
\cite{otsuka_band_2017,sawamura_nearest-neighbor_2017}.

In this paper, we evaluate $\epsilon_\infty$ not in bulk calculations
with such approximations used in Ref.\onlinecite{shishkin_accurate_2007},
but by slab models with finite electric bias voltage.
We treat five ionic materials, LiF, KF, NaCl, MgO and CaO.
We put a slab in the middle of vacuum region in a supercell.
The electric field is applied by the effective screening medium
(ESM) method given by Otani and Sugino \cite{otani_first-principles_2006}.
We obtain $\epsilon_\infty$ from the ratio of slopes of the electrostatic
fields at slab region and at vacuum region.
Our approach is based on the self-consistent method, thus we do not need to
utilize approximations as was used in Ref.\onlinecite{shishkin_accurate_2007}.
Since we explicitly treat the response to the bias,
our method includes higher-order effects in a self-consistent manner.

Our findings are that the calculated $\epsilon_\infty$ in QSGW for the
slab models are very close to experimental values.
This is in contrast to the fact that $\epsilon_\infty$ in RPA of QSGW
are generally $\sim$ 20 percent smaller than experimental values.
This indicates that the vertex correction at the level of derivative of the QSGW self-energy
should make $W$ be in agreements with experiments.
Our results is consistent with Table II in Ref.\onlinecite{shishkin_accurate_2007}.

We can interpret the enhancement of screening, represented by the enlargement of $\epsilon_\infty$,
as the size of the vertex correction for the proper polarization $P$.
%
Note that the vertex correction we evaluate is not what is defined in
the Hedin's equation \cite{hedin_effects_1969}.
In the equation, we see $P= -i G G \Gamma$, that is,
the vertex function $\Gamma$ is for the correction to $P= -i G G$, 
where $G$ denotes the Green's function.
Instead, we rather evaluate $\Gamma$ for $P=-i G_0 G_0 \Gamma$,
where $G_0$ is the bare Green's function. 
To clarify the above theoretical point on $\Gamma$, we give an extensive discussion in Sec.\ref{qpper}.
We explain role of $\Gamma$ in the two kinds of perturbation theories.
In Sec.\ref{QSGWESM}, we explain QSGW+ESM, an implementation of QSGW
combined with ESM. The QSGW+ESM for slab models should be very useful
not only for our purpose here, but also for others where usual GGA+ESM have difficulties.
In Sec.\ref{result}, we show our results of $\epsilon_\infty$.
Then they are interpreted as the vertex correction. In
Sec. \ref{subsec:rationale}, we give a
rationale of QSGW80, followed by a summary.



\section{QP-based perturbation vs. $G$-based perturbation}
\label{qpper}
To make our motivation in this paper clarified, we have to clarify
the difference between the quasiparticle-based perturbation (QbP) and
the Green's function-based perturbation (GbP). 
QbP is based on the Landau-Silin's QP theory,
while GbP is on the Hedin's one. 
To illustrate the difference between QbP and GbP,
we give a narrow-band model as follows. The model represent situations where we have good
QP picture (= independent-particle picture).

In advance, remind that we mainly have two kinds
of excitations in the paramagnetic electronic systems.
That is, the multi-particle excitations, 
and the collective excitation such as plasmons.
The former is described by QPs interacting each other.
Note that plasmons is located at high energy
because of the long-range Coulomb interaction \cite{pines66}.
These excitations can be hybridized. For example, we know pseudo plasmon
in Silver, where one-particle excitations of $3d$ electrons are hybridized with plasmons.

\subsection{narrow-band model to explain the QP-based perturbation}
\label{qpper1}
\begin{figure}[ht]
 \caption{\label{fig:qbp}
 Band structure of a narrow-band model to
 illustrate the quasiparticle based perturbation.
 CBM and VBM are the acronyms of conduction-band minimum and 
 valence-band maximum.
 The width of bands $B_{\rm C}$ and $B_{\rm V}$ are smaller than the band gap
 $E_{\rm G}$. As in the text, we expect well-defined QPs in this model. }
\includegraphics[width=10cm]{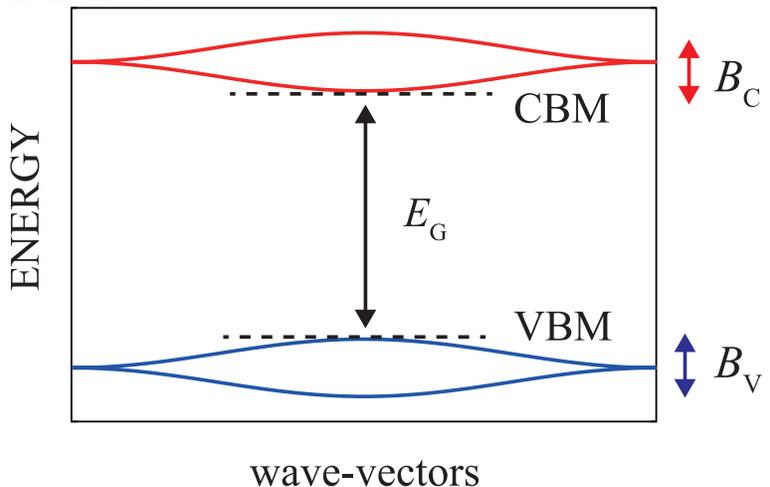}
\label{fig:QbP}
\end{figure}

The QPs based on the Landau-Silin's Fermi liquid theory is originally for
metals \cite{pines66}. However, the idea of QPs are
rather easily applicable to insulators. We can consider QbP based on the QPs.
To illustrate this, let us consider a narrow-band model, a paramagnetic case given by a
Hamiltonian $H$, which has an one-body term represented by
finite numbers of Wannier functions in the primitive cell,
and the Coulomb-like interaction $\frac{e^2}{\epsilon' |\bfr-\bfr'|}$,
where $\epsilon'$ is a constant.
We consider a case that it gives the QPs shown in Fig.\ref{fig:QbP}
given by the one-particle Hamiltonian $H_0$.

Based on the perturbation, we expect that the lifetimes of all the electrons are infinite
because the band gap $E_{\rm G}$ is large enough to forbid all the
electrons decaying into lower energy electrons accompanying electron-hole pairs, holes as well.
In other words, all bands are within the threshold of impact ionization \cite{kotani_impact_2010}.
Thus QPs described by $G_0=1/(\omega-H_0)$ should give well-defined
one-particle excitations of the narrow-band model.

This conclusion should be essentially kept even when we fully turn on the
interaction as long as the following conditions are well satisfied.
First, excitons, binding states of electron-hole pairs,
should be only slightly at lower energies than $E_{\rm G}$.
Second, plasmons should be located at high enough energies 
so that the plasmon are hardly hybridized with the one-particle excitations.
Under these assumptions, we have well-defined QPs.
We can consider a path of adiabatic connection with keeping the
well-defined QPs given by $H_0$ since the QP spectrum
is clearly separated from the other excitations.
It can be written as $H_\lambda =H_0+ (H_\lambda-H_0)$ for $\lambda$ is from
zero to unity, where $H=H_{\lambda=1}$. Note that $(H_\lambda-H_0)$ need to contain
$\lambda$-dependent one-body term.

In this narrow-band model, QbP should be suitable;
QPs are interacting each other by $(H_{\lambda=1}-H_0)$.
As for the proper polarization function, we can take all the
non-interacting two-body QP excitations correctly by $P_0= -i G_0 G_0$.
Based of this QbP, $G_0W_0$ approximation, $\Sigma=i G_0  W_0$ where $W_0=v/(1-vP_0)$,
is physically justified. That is, it describes how the motion of QPs in $G_0$ is perturbed by the
dynamical self-interaction given by $W_0$ in RPA.
Most of all first-principles calculations in literatures are implicitly
based on this QbP, while some latest literatures \cite{kutepov_electronic_2016,kutepov_self-consistent_2017}
are based on GbP explained in Sec.\ref{GbP}.
QSGW is a method to determine $H_0$ self-consistently in QbP.

\subsection{vertex function  $\Gamma_{\rm QbP}$ in the QP-based perturbation}
To improve $G_0W_0$, we may include electron-hole correlation.
The corrections replace $P_0= -i G_0 G_0$ with $P$ where
we include the correlation via the Bethe-Salpeter equation (ladder
diagrams). That is, we include two-body spectrum in the proper polarization
accurately. This lets us to use $W=1/(1-vP)$ instead of $W_0$, resulting the $G_0 W$
approximation as $\Sigma=i G_0 W$.
In this paper, we concentrate on the $G_0 W$ approximation
as in the case of sc$GW$(e-h) in Ref.\onlinecite{shishkin_accurate_2007}.
In QbP, we thus define the vertex function $\Gamma_{\rm QbP}$ for $W$
as $P= -i G_0 G_0 \Gamma_{\rm QbP}$. Roughly speaking, $P=P_0 \Gamma_{\rm QbP}$.
We give the numerical evaluation for this $\Gamma_{\rm QbP}$ via the
evaluation of $\epsilon_\infty$ as shown in Sec.\ref{result1}.

To go beyond $\Sigma=i G_0 W$ approximation here, 
we need to take into account three-particle intermediate states.
However, it is not theoretically straightforward because of
a double counting problem that $\Sigma=i G_0 W$
already partially takes into account such states.
We will need to construct theories of
three-particle problem without double counting
along the line of first-principles calculations.
We do not treat this problem in this paper.


\subsection{$G$-based perturbation and vertex function $\Gamma_{\rm GbP}$}
\label{GbP}
Let us consider how we can apply GbP to the narrow-band model.
In contrast to $G_0$ in QbP, the one-body Green function $G$ has
complex meanings. We have imaginary part of $G$ at high energies,
representing QPs hybridized with plasmons (plasmarons).
Because of sum rule for ${\rm Im}[G]$, the QP parts are suppressed by $Z$-factor.
Thus there is a problem that $P= -i G G$ do not contain the two-body non-interacting
excitations with the correct weight, in contrast to the case $P_0=-i G_0 G_0$.

In principle, this problem is corrected by including the vertex function
$\Gamma_{\rm GbP}$ in the Hedin's equation 
to determine the one-particle Green's function $G(1,2)$ \cite{hedin_effects_1969}.
Because Hedin's equation is theoretically rigorous, we expect
$-i G G \Gamma_{\rm GbP} \approx P_0=-i G_0 G_0$ in the model, under the
discussion of Sec.\ref{qpper1} that $P_0$ gives good approximation for the model.
That is, contributions related to the collective excitations and
renormalization factors $Z$ in $-i G G$ should be virtually taken away
by the factor $\Gamma_{\rm GbP}$.
However, such numerical calculations should be computationally very
demanding \cite{Bechstedt97}.
Similar discussion of $Z$-factor cancellation is also seen when we
multiply $G_0$ to $W$. That is,
we should have $G_0 W \approx G W \Gamma_{\rm GbP}$ since QbP correctly treats the model. 
Our analysis here is consistent with Takada's analysis
based on the Ward identity \cite{takada2016}.

QbP should be generally superior to GbP even in real materials.
In contrast to GbP, QbP is quite simple and physically convincing.
We should not be confused with the similarity of QbP and GbP.
In this paper, we evaluate $\Gamma_{\rm QbP}$.
In the following, we calculate the enhancement of the screening effect.
Then we evaluate the size of the ratio $P/P_0$
from the comparison between calculated $\epsilon_\infty$ in RPA and $\epsilon_\infty$
in the slab models. This ratio gives the size of $\Gamma_{\rm QbP}$.


\section{QSGW combined with Effective screening medium method}
\label{QSGWESM}
To calculate $\epsilon_\infty$, we put a slab
in the middle of vacuum region in a supercell.
$\epsilon_\infty$ is calculated from the difference of slopes
in the vacuum region and in the slab region under small bias voltage.
The supercell we use are detailed at the beginning of Sec.\ref{result}.
In such calculations, we can obtain $\epsilon_\infty$ beyond the bulk
calculation in RPA as we explain in the next paragraph.
That is, we can obtain $\epsilon_\infty$ including the effect of the vertex correction.



To explain how the effect is included, let us consider slab calculations in the case of GGA at first.
We first perform self-consistent calculation under zero bias.
Then we perform self-consistent calculation under the finite bias
(theoretically, it should be infinitesimally small).
Then we have the difference of the electron density
$\delta n(\bfr)$ between the two calculations.
Simultaneously, we have a corresponding response of the one-particle
potential given as
$\delta V(\bfr) = \int d^3r' v(\bfr-\bfr') \delta n(\bfr') +
\frac{\partial V^{\rm GGA}_{\rm xc}}{\partial n(\bfr)} \delta n(\bfr)$.
The last term is the difference in the exchange-correlation (xc) potential
caused by $\delta n(\bfr)$ self-consistently.
That is, the derivative $\frac{\delta V(\bfr)}{\delta n(\bfr')}$ contains
the contribution of the xc kernel $f_{\rm xc}=\frac{\partial V^{\rm GGA}_{\rm xc}}{\partial n(\bfr)}$.
Under the bias, we can obtain $\epsilon_\infty$, from the ratio of
slopes of the electrostatic potential in the vacuum region and in the slab region.
It should contain the contribution from $f_{\rm xc}$, which is identified as the
vertex correction in GGA.

This is essentially the same in QSGW.
Recall that the self-energy in QSGW denoted as
$V^{\rm QSGW}_{\rm xc}(\bfr,\bfr')$ is a static non-local potential,
replacing $V^{\rm GGA}_{\rm xc}$.
The derivative of the one-particle potential is given as
$\delta V(\bfr,\bfr') = \int d^3r'' v(\bfr-\bfr'') \delta n(\bfr'') +
 \delta V^{\rm QSGW}_{\rm xc}(\bfr,\bfr')$, where the last term play
a role of $\frac{\partial V^{\rm GGA}_{\rm xc}}{\partial n(\bfr)} \delta n(\bfr)$.
Note that $\delta V^{\rm QSGW}_{\rm xc}(\bfr,\bfr')$ is determined
self-consistently, although it is not so simply given as
$\frac{\partial V^{\rm QSGW}_{\rm xc}}{\partial n(\bfr)} \delta n(\bfr)$.
Our calculations include the contribution of $\delta V^{\rm QSGW}_{\rm xc}(\bfr,\bfr')$
self-consistently as in the case of GGA.
Our method is on the same spirit of solving
the Bethe-Salpeter equation in Ref.\onlinecite{PhysRevLett.122.237402}.

Our QSGW+ESM is implemented in a first-principles package {\tt ecalj}
\cite{deguchi_accurate_2016,ecalj} which is based on a mixed-basis
method, the augmented plane wave (APW) and Muffin-tin (MT) orbital method (the PMT method)
\cite{pmt1,kotani_linearized_2013,kotani_quasiparticle_2014,kotani_formulation_2015}.
The PMT method is an all-electron full potential method which uses not
only the APW basis in the LAPW method, but also the MT orbitals in
the LMTO method simultaneously in the expansion of eigenfunctions.
It also use the local orbital basis \cite{PhysRevB.43.6388}.
On top of the PMT method, we had implemented the QSGW method \cite{kotani_quasiparticle_2014,deguchi_accurate_2016}.
In PMT, we use very localized untuned MTOs which contains damping factor
$\exp(- \kappa r)$, where $\kappa$ are fixed
to be 1/bohr and/or 2/bohr, together with low-cutoff APWs ($\leq$3 Ry).
We do not need empty spheres since the APWs can handle vacuum regions of slab models.
The charge density is represented in the three component representation,
'smooth part','true part within MT', and 'counter part within MT' as in
the case of PAW method \cite{PAW}.
In contrast to the other $GW$ methods which requires the
Wannier-interpolation technique to make band plots in
the whole Brillouin zone, we can make band plots easily without
resorting to the technique \cite{deguchi_accurate_2016}.
In the following, we show how to implement ESM in
the PMT method, after an explanation of general theory of ESM.

\subsection{The electrostatic potential in the Effective screening medium method}
We apply the ESM method \cite{otani_first-principles_2006} to slab
models under an external electric field.
We treat a supercell with periodic boundary condition
where we have a slab with periodicity in the {\it xy}-plane. The slab is at the middle of supercell.
Position in the cell is specified by $\bfr=(\bfr_{//},z)$. 
Planes at $z=-z_0$ and at $z=z_0$ are the left and right ends of the
supercells. The electrostatic potential is calculated from the charge
density in the supercell assuming two electrodes are at $z=\pm z_0$
(we set $z_1=z_0$ in Fig.1 of Ref.\onlinecite{otani_first-principles_2006}.)
for applying voltage to the supercell.
As we summarize as follows, the ESM in DFT is formulated from the total energy minimization,
however, it is not true in QSGW since QSGW itself is not formulated from
the total energy minimization. After we obtain the following key equation
\req{eq:vone} to determine electrostatic potential.
We use it even in QSGW.


Let us start from the energy functional of DFT in the ESM. It is written as
\begin{eqnarray}
E[n] = E^{\rm kin}[n] +E^{\rm xc}[n] + E^{\rm es}[n] + E^{\rm app}[n].
\label{elda}
\end{eqnarray}
Here, we have kinetic energy $ E^{\rm kin}[n]$, xc
energy $E^{\rm xc}[n]$, and the electrostatic energy $E^{\rm es}[n]$ terms. In addition,
the last term is the applied electrostatic
term $E^{\rm app}[n]=\int d^3r V^{\rm app}(\bfr) (n(\bfr)+n_{\rm N}(\bfr))$,
where $n(\bfr)$ and $n_{\rm N}(\bfr)$ are the electron density
and the the charge density of nuclei, respectively; 
$V^{\rm app}(\bfr)$ is a linear function of $z$, representing the external field.

In ESM, we enforce the periodicity in the supercell for the electrostatic potential.
Thus we use $V^{\rm app}(\bfr)s(\bfr)$ instead of 
$V^{\rm app}(\bfr)$, where we introduce a support function $s(\bfr)$ 
which is unity for most of all regions, but is going to be zero at
$z=-z_0$ and $z=z_0$. It is different from unity only near the boundaries, 
$z\approx -z_0$ or $z\approx z_0$.
Thus the potential $V^{\rm app}(\bfr)s(\bfr)$ recover the periodicity
of the supercell. A constant can be added to $V^{\rm app}(\bfr)$
so that it keep smooth periodicity over $z=\pm z_0$.
As long as we use large enough vacuum region, we
have little electrons near the boundaries.
Thus the choice of $s(\bfr)$ is irrelevant.

A key in ESM is that we use the Green function $\bar{v}(\bfr,\bfr')$ for the
electrostatic energy $E^{\rm es}[n]$ instead of the Coulomb interaction
$v(\bfr-\bfr')$ in usual the GGA calculations.
As in Ref.\onlinecite{otani_first-principles_2006},
$\bar{v}(\bfr,\bfr')$ contains not only the Coulomb interaction $v(\bfr-\bfr')$ but
also the effects due to the polarization of virtual electrodes,
which are at $z=\pm z_1$ (we use $z_0=z_1$ in our calculations here).
Polarization of the slab occurs with keeping the electrostatic potential being constants at electrodes.
Corresponding to $V^{\rm app}(\bfr)s(\bfr)$,
we use $s(\bfr)\bar{v}(\bfr,\bfr')s(\bfr')$ instead of $\bar{v}(\bfr,\bfr')$ in practice. 
Then we have well-defined Kohn-Sham total energy with keeping the
periodic boundary condition for given $V^{\rm app}(\bfr)$.

The minimization of $E[n]$ with respect to $n(\bfr)$ gives the
Kohn-Sham potential $V(\bfr)$ as
\begin{eqnarray}
V(\bfr)= \int d^3r' \bar{v}(\bfr,\bfr') (n(\bfr')+n_{\rm N}(\bfr')) +  V^{\rm app}(\bfr) + V^{\rm xc}(\bfr),
\label{eq:vone}
\end{eqnarray}
Hereafter, we skip $s(\bfr)$ for simplicity.

In QSGW \cite{kotani_quasiparticle_2014}, we cannot derive its
fundamental equation from the energy minimization.
Thus the formulation of QSGW+ESM is not exactly along the line above.
However, we can use the one-particle potential of \req{eq:vone} in the
self-consistent cycle, where $V^{\rm xc}(\bfr)$ is replaced by a
static version of the self-energy \cite{Faleev04}.
Thus, in principle, it is straightforward to perform QSGW+ESM.\\

\subsection{ESM in the PMT method}
In the PMT method, electron density (and also the charge density) 
is represented by the three component formalism described in
Ref.\onlinecite{kotani_formulation_2015},
originally introduced by Soler and Williams \cite{soler89,soler90,soler93}.
At first, space is divided into MT regions and interstitial regions.
Then electron density is represented by three components as
$n=\{n_0(\bfr),\{n_{1,a}(\bfr)\},\{n_{2,a}(\bfr)\} \}$ where $a$ is the
index of atomic sites in the primitive cell. 
Following Ref.\onlinecite{kotani_formulation_2015},
this is simply expressed as $n=n_0~\oplus n_1~\ominus~n_2$.
The 0th component $n_0(\bfr)$ is the spatially smooth functions,
expanded in analytic functions, that is, planewaves, Gaussians, and smooth Hankel functions
\cite{kotani_quasiparticle_2014}.
The 1st components $n_{1,a}(\bfr)$
is the true electron density within MT at $\bfR_a$.
The 2nd components $n_{2,a}(\bfr)$ is the counter part,
that is, the projection of $n_0(\bfr)$ into the MT at $\bfR_a$.
$n_0(\bfr)$ and $n_{2,a}(\bfr)$  are identical within MT at $\bfR_a$ up
to given angular momentum cutoff in their spherical harmonics expansion.

We can get all charge density 
$n^{\rm Zcv}=n^{\rm Zcv}_0~\oplus n^{\rm Zcv}_1~\ominus~n^{\rm Zcv}_2$
by adding the ion-core contribution to $n$.
Then we apply the multipole transformation clearly defined in Ref.
\onlinecite{kotani_formulation_2015}, resulting 
$\bar{n}^{\rm Zcv}_0 \oplus \bar{n}^{\rm Zcv}_1 \ominus
\bar{n}^{\rm Zcv}_2$ as shown in Eq.(28-30) in Ref.\onlinecite{kotani_formulation_2015}.
The transformation makes
$\bar{n}^{\rm Zcv}_0(\bfr)$, $\bar{n}^{\rm Zcv}_{1,a}(\bfr)$, and $\bar{n}^{\rm Zcv}_{2,a}(\bfr)$ 
have the same multipole in each MT site at $\bfR_a$, although physically observable density unchanged.
The 1st components $\bar{n}^{\rm Zcv}_{1,a}(\bfr)$, unchanged by the transformation, are  
the sum of ion-core charge density and $n_{1,a}(\bfr)$.

From the smooth density $\bar{n}^{\rm Zcv}_0(\bfr)$, we can calculate
electrostatic potential as
$V^{\rm es}_0(\bfr)= \int d^3r' \bar{v}(\bfr,\bfr') \bar{n}^{\rm
Zcv}_0(\bfr') +  V^{\rm app}(\bfr)$. 
This gives correct interstitial part of the potential $V^{\rm es}_0(\bfr)$
calculated from all charge density. The values of $V^{\rm es}_0(\bfr)$
at MT boundaries are used to determine the electrostatic potential within MTs.

We can use usual procedure to determine the electrostatic potential within MTs.
In each MT, we have 1st and 2nd components
$\bar{n}^{\rm Zcv}_{1,a}(\bfr)$ and $\bar{n}^{\rm Zcv}_{2,a}(\bfr)$, which have the same multipole. 
With the condition that the electrostatic potential is zero at the MT
boundary, we can calculate the potential generated by the difference of
the 1st and 2nd components.

Thus we finally have the electrostatic potential $V^{\rm es}(\bfr)$ represented in the three component formalism.
With this potential, we can perform self-consistent calculations for slab models.




\section{Results}
\label{result}
\subsection{Optical dielectric constants via the slab model}
\label{result1}
\begin{figure}[ht]
 \caption{\label{fig:esm}
 A slab (18 atoms per cell) is placed in in the middle of a supercell
(60 a.u. width along the $z$ axis which is perpendicular to the slab),
with electrodes at the left and right ends.
At the top panel, we show $V_z^{\rm es}(z,E)$ for $E=0.2$ Ry and $E=0.0$ Ry.
In the bottom panel, we show their difference $\Delta V_z^{\rm es}(z)$.
From the ratio of two slopes of $\Delta V_z^{\rm es}$ 
in the slab region(green) and in the vacuum region(violet), we obtain $\epsilon_\infty({\rm Slab})$.
We have better numerical accuracy by using $\Delta V_z^{\rm es}(z)$
instead of $V_z^{\rm es}(z,0.2 {\rm  Ry})$ directly.
}
\includegraphics[width=8cm]{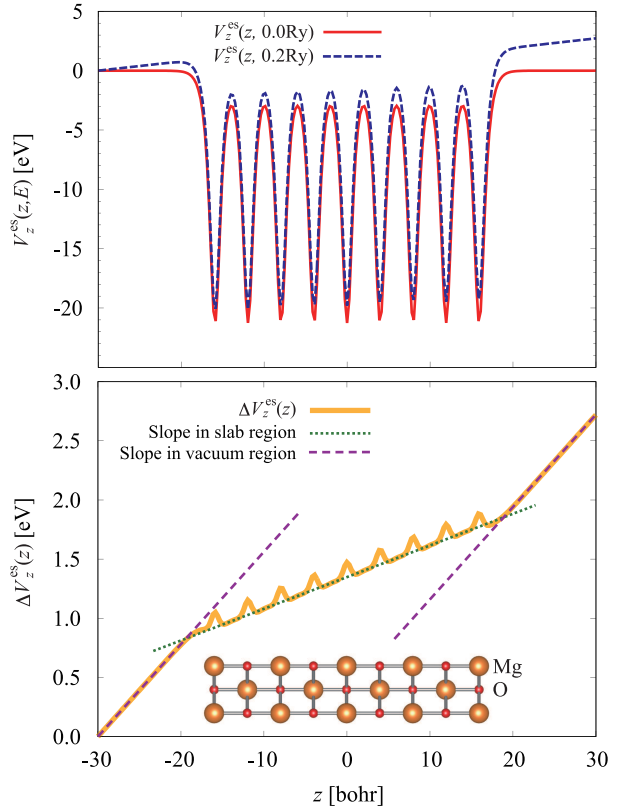}
\end{figure}

In Fig.\ref{fig:esm}, we illustrate our treatments 
in the slab models for five NaCl-structure ionic materials, 
where we use $\pm z_0=\pm 30$ a.u.
We use slabs made of nine layers, 18 atoms in the supercell.
We use experimental lattice constants of bulk materials, without relaxation of atomic positions.
The electrostatic potential $V_z^{\rm es}(z,E)$ are
the average of $V_0^{\rm es}(\bfr)$ in the {\it xy} plane under the bias voltage $E$.
We plot the cases of $E=0.2$ Ry and of $E=0.0$ Ry.
We show $\Delta V_z^{\rm es}(z)=V_z^{\rm es}(z,0.2 {\rm Ry})-V_z^{\rm es}(z,0.0 {\rm Ry})$
in the bottom panel in Fig.\ref{fig:esm}. 
$\Delta V_z^{\rm es}(z,E)$ changes linearly as a function of $z$ in the vacuum
regions and in the slab regions.
From the ratio of two slopes of $\Delta V_z^{\rm es}$ 
in the slab region and in the vacuum region, we obtain
$\epsilon_\infty({\rm Slab})$.
\begin{table}[h]
\caption{Calculated optical dielectric constant
 $\epsilon_\infty$. `RPA' are in bulk calculations with
 local field correction (LFC). `RPA(noLFC)' are without LFC.
 `Slab' are calculated from the slab models in the setting of
 Fig.\ref{fig:esm}. Ratios $\eta=\frac{\epsilon_\infty({\rm RPA})}{\epsilon_\infty({\rm Slab})}$
 and $\gamma=\frac{\epsilon_\infty({\rm Slab})-1}{\epsilon_\infty({\rm RPA})-1}$ are calculated just simply
 from the values of $\epsilon({\rm QSGW,RRA})$ and $\epsilon({\rm QSGW,Slab})$. }
\label{table:epstab}
\begin{tabular}{l| c c c c c c c}
\hline
\hline
& & RPA      & RPA     &  Slab &  $\eta$ &  $\gamma$  & Experiments \\
& & (noLFC)  &         &       &           &             & \cite{alkali,mgo-exp,cao-exp}  \\
\hline
\multirow{2}{*}{LiF}
    &GGA  &  2.04 & 1.95  & 2.01 & & &\multirow{2}{*}{1.96}\\
    &QSGW &  1.73 & 1.67  & 1.94 & 0.86 &1.40\\
\hline 
\multirow{2}{*}{KF}  &GGA  &  2.16  & 1.96   & 1.94 & & &\multirow{2}{*}{1.85}\\
    &QSGW &  1.79 & 1.68  & 1.86 & 0.90 & 1.26\\ 
\hline 
\multirow{2}{*}{NaCl}&GGA  &  2.70 & 2.33  & 2.42 & & &\multirow{2}{*}{2.34}\\
    &QSGW &  2.13 & 1.92  & 2.31 & 0.83 & 1.42\\ 
\hline 
\multirow{2}{*}{MgO} &GGA  &  3.17 & 2.96  & 3.09 & & &\multirow{2}{*}{2.96}\\
    &QSGW &  2.50 & 2.37  & 2.91 & 0.81 &1.39\\ 
\hline 
\multirow{2}{*}{CaO} &GGA  &  3.94 & 3.59  & 3.68 & & &\multirow{2}{*}{3.33}\\
    &QSGW &  2.88 & 2.68  & 3.31 & 0.81 &1.38\\ 
\hline
\hline
\end{tabular}
\end{table}

Our main results are $\epsilon_\infty$ calculated from slab models in QSGW,
$\epsilon_\infty({\rm QSGW,Slab})$, in Table \ref{table:epstab}.
Note that $\epsilon_\infty({\rm Slab})$ contains the effect of vertex
corrections based on QbP (See Sec.\ref{qpper}), because changes of the self-energy caused by the bias $E$
are self-consistently taken into account.
Numerical reliability of our calculations
are estimated to be $\lesssim 1$ percent.
See supplemental materials for computational details \cite{supply}.
In Table \ref{table:epstab}, we also show bulk values
$\epsilon_\infty({\rm RPA})$. To obtain them, we first perform
self-consistent calculations in QSGW for bulk materials.
Then we calculate $\epsilon_\infty$ in the random-phase
approximation (RPA) with/without local field correction (LFC).
We also show $\epsilon_\infty$ in GGA together.



The QSGW values are in good agreements with experiments.
For example, $\epsilon_\infty({\rm QSGW,Slab})$=1.94 for LiF gives surprisingly
good agreement with $\epsilon_\infty({\rm Experiment})$=1.96. \hspace{20pt}
In contrast, $\epsilon_\infty({\rm QSGW,RPA})$=1.67 is very smaller than $\epsilon_\infty({\rm QSGW,Slab})$=1.94.
These are generally true in all other materials.
We see that ratios $\eta=\epsilon_\infty({\rm QSGW,RPA})/\epsilon_\infty({\rm QSGW,Slab})$
in Table \ref{table:epstab} are $\sim$0.8. 
This is consistent with Ref.\onlinecite{kotani07a}
where $\epsilon_\infty$ for ZnO, Cu$_2$O, MnO, and NiO are presented.
From a point of view to estimate the enhancement factors
($\approx$ vertex $\Gamma$ ) of the proper polarization, we may consider ratios
$\gamma=\frac{\epsilon_\infty({\rm Slab})-1}{\epsilon_\infty({\rm RPA})-1}$.
As shown in Table \ref{table:epstab}, $\gamma\sim$1.4. 
Since $\epsilon_\infty({\rm QSGW,Slab})$ gives very good agreements with $\epsilon_\infty({\rm experiment})$,
we can say that the vertex correction for bulk should give the
difference between $\epsilon_\infty({\rm RPA})$ and $\epsilon_\infty({\rm experiment})$ very well, where
the vertex correction is calculated at the level of the functional
derivative of the self-energy in QSGW; See Sec.\ref{QSGWESM}.

This is in contrast to the case of GGA. For example, look into the case
of LiF. The difference $\epsilon_\infty({\rm GGA,Slab})-\epsilon_\infty({\rm GGA,RPA})=2.01-1.95=0.06$ is
very small. The difference is originated from the xc kernel $f_{\rm xc}$ in the density
functional perturbation theory. This is consistent with results in Ref.\onlinecite{botti_long-range_2004} 
where they explicitly evaluate $f_{\rm xc}$ in GGA for bulk materials.
Note that $\epsilon_\infty({\rm GGA,Slab})=2.01$ is a little larger than
$\epsilon_\infty({\rm experiment})=1.96$; this is true for all other materials.
We see that the contributions of vertex corrections $f_{\rm xc}$ do not necessarily improve
agreements; $\epsilon_\infty({\rm GGA,Slab})$ give poorer
agreement with $\epsilon_\infty({\rm experiment})$ than $\epsilon_\infty({\rm GGA,RPA})$.




\subsection{Rationale for QSGW80}\label{subsec:rationale}
\begin{table}[ht]
\caption{Calculated band gaps (eV) of bulk materials. 
 In QSGW80, we show self-consistent results with the hybrid
 xc potential, 80 \% QSGW+ 20 \% GGA.
 QSGW80nosc specifies one-shot calculations with the hybrid potentials
 after QSGW 100\% self-consistent calculations. QSGW80nosc is slightly
 larger because it is not fully self-consistent under such xc potential.
 See  Ref.\onlinecite{deguchi_accurate_2016}.
}
\label{table:gaptab}
\begin{tabular}{l|c|ccccc}
\hline
\hline
    &  Experiments  & \multirow{2}{*}{QSGW}  & \multirow{2}{*}{QSGW80} & \multirow{2}{*}{QSGW80nosc} & \multirow{2}{*}{GGA}  \\
    &  \cite{lif-gap,kf-nacl-gap,mgo-gap,cao-gap} &  &  &  &  \\
\hline
 LiF &   13.6 & 16.04   & 14.53 & 14.85 & 9.52 \\
 KF  &   10.9 & 11.78 & 10.53 & 10.82 & 6.43   \\
 NaCl&    8.6 & 9.51 & 8.55 & 8.76 & 5.37 \\
 MgO &   7.77 &  8.86  & 7.91 & 8.10 & 4.86\\
 CaO &    7.1 & 7.45 & 6.57  & 6.74 & 3.69 \\
 \hline
\end{tabular}
\end{table}

Our result in Sec.\ref{result1} shows that the vertex correction
should be included in the proper polarization $P$ to obtain
$\epsilon_\infty$ in agreement with experiments.
We have to use such $P$ in the QSGW self-consistent cycle.
Such improved QSGW self-consistency can be identified as a
self-consistent method in the $G_0W$ approximation on the basis of QbP.
Ref.\onlinecite{shishkin_accurate_2007} by Shishkin {\it et al.} gives a method on this idea.

However,
their method is too expensive for computational efforts to apply wide-range of materials.
In fact, although their method was applied to calculate ionization potentials
in Ref.\onlinecite{PhysRevLett.112.096401}, it was not
so satisfactory because calculations are performed in the combination of
simple materials (bulk calculations) with the supercell calculations in GGA.
Furthermore, no papers available to treat transition-metal oxides such as LaMnO$_3$ in their method.
We have to develop such an improved QSGW method applicable to wide range of materials.
Two requirements are the computational efficiency and the theoretical validity.

As a possibility to respect the efficiency, 
we can consider a hybridization method between QSGW and the density functional xc \cite{chantis06a}.
In QSGW80, a simple hybridization, 80 \% QSGW+ 20 \% GGA, we can see that it
works well for wide range of materials. 
Our present results support the method of QSGW80, which takes only the
80 percent of QSGW self-energy. We can identify QSGW80 as a simplification
of the method in Ref.\onlinecite{shishkin_accurate_2007}.
Ref.\onlinecite{PhysRevB.92.041115} also presents one another approximation at the level
of QSGW80 for the vertex correction in QSGW, resulting in similar good agreement with experiments.

Let us examine how QSGW80 is justified for materials calculated here.
This is by the fact that
$\eta=\epsilon_\infty({\rm QSGW,RPA})/\epsilon_\infty({\rm QSGW,Slab})$ in Table \ref{table:epstab}
are approximately 80 \%, and show {\it little material-dependency}.
Thus we expect that QSGW80 can mimic QSGW with the vertex corrections;
too large screened-exchange effect is reduced by the factor 0.8, with
adding 0.2 GGA term so as to keep the total size of the xc term.
In Table \ref{table:gaptab}, we show band gaps in QSGW and QSGW80 for materials
treated here. The band gaps are systematically too large in QSGW
in comparison with experimental values \cite{deguchi_accurate_2016},
while QSGW80 gives rather better agreements with experimental values.
In Ref.\onlinecite{deguchi_accurate_2016}, we checked the performance of the QSGW80 for ranges of materials.
As in the case of Ref.\onlinecite{shishkin_accurate_2007}, QSGW80 is
theoretically reasonable in the sense that the band gaps are improved by using the corrected $W$.
To go beyond QSGW80, we have to develop methods to take the vertex
correction into $W$ as was done in Ref.\onlinecite{shishkin_accurate_2007} in a simple manner.
Considering the fact that QSGW80 works well as shown in Ref.\onlinecite{deguchi_accurate_2016},
we may expect simple methods to represent the vertex correction
by a scalar factor or by limited number of parameters.
As long as we know, the vertex correction can be relatively insensitive
to materials, thus we expect some simple method might be available.

\section{Summary}
To clarify the importance of quasiparticle self-consistency in QSGW,
we have explained the quasiparticle based perturbation in Sec.\ref{qpper}.
Then we emphasize the importance of the self-consistency in the $G_0 W$
approximation. Then we obtain the quasiparticles (independent-particle) given by
$H_0$, and the interaction between the quasiparticles given by $W$.
The vertex correction in QbP is introduced.

We have performed QSGW calculations for slab models under electric field by means
of the ESM method. The calculated $\epsilon_\infty$ are in good agreements with experimental values.
Compared with $\epsilon_\infty$ in bulk calculation in RPA,
we evaluated the size of vertex corrections
as the functional derivative of the static self-energy in QSGW.
Our results on $\epsilon_\infty$ give a support to the method
by Shishkin, Marsman and Kresse \cite{shishkin_accurate_2007}.
As a simplified substitution of their method, we examined the performance of
QSGW80 \cite{deguchi_accurate_2016} for materials treated here.
The method QSGW+ESM developed for the calculations should be useful
even for other purposes such as bias-dependent
spin susceptibility in material theory, as well as practical device
applications and materials designs.

To go beyond usual QSGW, we should develop improved QSGW method
in the $G_0 W$ approximation, whereas we should use accurate $W$ by including the vertex correction.
Then we have virtually best division of $H=H_0+(H-H_0)$ where $H_0$
gives the optimum independent particle picture.

\begin{acknowledgments}
T. K. thanks to supporting by JSPS KAKENHI Grant Number 17K05499.
We also thank the computing time provided by Research Institute for Informat\
ion Technology (Kyushu University).
H. S. thanks to the computing resources provided by
 the supercomputer system in RIKEN (HOKUSAI) and the supercomputer system in ISSP (sekirei).
\end{acknowledgments}

\bibliography{refkotani}
\end{document}